\documentstyle[prl,aps,epsf,floats]{revtex}
\begin{document}
\draft
\twocolumn[\hsize\textwidth\columnwidth\hsize\csname @twocolumnfalse\endcsname
\title{Short-range spin correlations and pseudogap in underdoped cuprates}   
\author{Bumsoo Kyung}
\address{D\'{e}partement de physique and Centre de recherche 
sur les propri\'{e}t\'{e}s \'{e}lectroniques \\
de mat\'{e}riaux avanc\'{e}s. 
Universit\'{e} de Sherbrooke, Sherbrooke, Qu\'{e}bec, Canada J1K 2R1}
\date{June 22, 2001}
\maketitle
\begin{abstract}

    In this paper we show that 
local spin-singlet amplitude with $d$-wave symmetry
can be induced by short-range spin correlations
even in the absence of pairing interactions.
In the present scenario for the pseudogap,
the normal state pseudogap is caused by the induced local
spin-singlet amplitude due to short-range spin correlations,
which compete in the low energy sector with superconducting
correlations to make $T_{c}$ go to zero near half-filling.
\end{abstract}
\vskip2pc]
\narrowtext

   The recent discovery of pseudogap in underdoped high  
$T_{c}$ cuprates has challenged condensed matter physicists for several  
years. 
The pseudogap behavior\cite{Timusk:1999} is observed as strong suppression 
of low frequency spectral weight below some characteristic temperature
$T^{*}$ higher than transition temperature $T_{c}$.
This anomalous phenomenon has been observed in 
angle resolved photoemission spectroscopy
(ARPES),\cite{Ding:1996,Loeser:1996}
specific heat,\cite{Loram:1993}
tunneling,\cite{Renner:1998}
NMR,\cite{Takigawa:1991} and
optical conductivity.\cite{Homes:1993}
One of the most puzzling questions in pseudogap phenomena is why 
$T^{*}$ has a completely different doping dependence from $T_{c}$,
in spite of their possibly close relation.
In this paper we demonstrate that
induced local spin-singlet amplitude due to short-range spin correlations
can cause a normal state pseudogap with $d$-wave symmetry
even in the absence of pairing interactions.

   First of all we argue that there are 
two energy scales in the problem, 
because the pseudogap appears as a crossover phenomenon according to
experiments.
The low energy (or long distance) physics of antiferromagnetic (AF) 
and superconducting (SC) correlations is well captured
by a {\it static} mean-field approach, while the relatively high
energy (or short distance) physics of the pseudogap is 
invisible in such a study.
Thus we resort to fluctuation theory in order to describe
the dynamical nature of the pseudogap, and to determine
$T^{*}$ and pseudogap size $\Delta_{pg}$.
The mean-field result of the $t-J$ model
will be used below solely
to find the onset of
leading correlations, and to compute mean-field AF and SC  
order parameters for the calculation of local spin-singlet
amplitude.

\begin{figure}
 \vbox to 6.5cm {\vss\hbox to -5.0cm
 {\hss\
       {\includegraphics{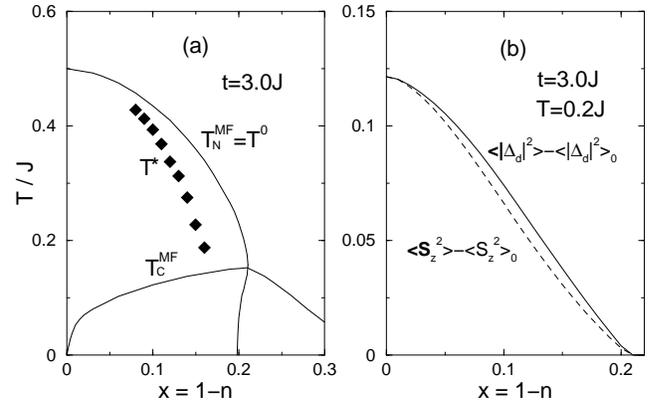}
       }
  \hss}
 }
\caption{(a) Calculated mean-field phase diagram in doping ($x=1-n$)
         and temperature ($T$) plane in the $t-J$ model for $t/J=3.0$.
         $T^{MF}_{N}$ and $T^{MF}_{c}$ are mean-field AF and SC
         ordering temperatures, respectively.
         The filled diamonds are the pseudogap
         temperature determined from the
         single particle spectral function.
         (b) Interaction induced local spin-singlet (solid curve) and spin
         (dashed curve) amplitudes for $t/J=3.0$ and $T=0.2J$.}
\label{fig1}
\end{figure}

   The mean-field $t-J$ Hamiltonian reads
\begin{eqnarray}
 H_{MF} = \sum_{\vec{k}, \sigma}\varepsilon(\vec{k})
                  c^{\dag}_{\vec{k},\sigma}
                  c_{\vec{k},\sigma}
         -4Jm\hat{m} -Js(\hat{s}+\hat{s}^{\dag}) \; ,
                                                           \label{eq10}
\end{eqnarray}
where
$\varepsilon(\vec{k}) \simeq -2tx(\cos k_{x}+\cos k_{y})-\mu$
with $x$ the hole density,
$\hat{m}=1/(2N)\sum_{\vec{k},\sigma}\sigma c^{\dag}_{\vec{k}+\vec{Q},\sigma}
                                         c       _{\vec{k},\sigma}$
and 
$\hat{s}=1/N\sum_{\vec{k}}\phi_{d}(\vec{k})
   c_{\vec{k},\uparrow}  c_{-\vec{k},\downarrow}$
with $N$ the total number of lattice sites.
$m$, $s$ are mean-field AF, SC order parameters determined 
from $m=\langle \hat{m} \rangle$ and 
     $s=\langle \hat{s} \rangle$, respectively.
$\phi_{d}(\vec{k}) = \cos k_{x}-\cos k_{y}$ and 
$\vec{Q}$ is the (commensurate) AF wave vector $(\pi,\pi)$ in two dimensions.
In this paper we restrict ourselves to a uniform solution which is just enough
for our purpose.
In a mean-field approximation, mean-field order
already sets in when
the correlation length reaches roughly one
lattice spacing.
This forces the above mean-field phase line (Fig.~\ref{fig1}(a))
to be interpreted as the onset
of the corresponding short-range correlations.
We identify $T_{N}^{MF}$ with another crossover temperature
$T^{0}$ at which some magnetic experiments such as Knight shift 
show their maximum.
For the parameter ($t/J=3.0$) used in this paper, short-range spin  
correlations
disappear at $x=x_{c} \simeq 0.19-0.20$ at low temperature.

   We introduce spin-singlet\cite{Comment15} correlation function
\begin{eqnarray}
\chi_{pp}(i,\tau)  =  \langle T_{\tau} \Delta_{d}(i,\tau)
                                     \Delta^{\dag}_{d}(0,0) \rangle  \; ,
                                                           \label{eq20}
\end{eqnarray}
where 
$ \Delta_{d}(i) = \frac{1}{2}\sum_{\delta}g(\delta)
        (c_{i+\delta,\uparrow}        c_{i,\downarrow}
        -c_{i+\delta,\downarrow}      c_{i,\uparrow}) $ with 
$g(\delta) =  1/2$ for $\delta=(\pm 1,0)$, 
             -1/2  for $\delta=(0,\pm 1)$, and 0 otherwise.  
The spin-singlet correlation function is related 
to the {\em local} spin-singlet
amplitude through the sum rule
\begin{eqnarray}
   \frac{T}{N}\sum_{q}\chi_{pp}(q)e^{-i\nu_{m}0^{-}}
=  \langle  |\Delta_{d}(0)|^{2} \rangle
                                               \; ,
                                                           \label{eq40}
\end{eqnarray}
where $q=(\vec{q},i\nu_{m})$ and $\nu_{m}$ is bosonic Matsubara
frequencies.
$T$ is absolute temperature. 
In terms of renormalized vertex $V_{pp}$,\cite{Vilk:1997}
we approximate the spin-singlet correlation function  
\begin{eqnarray}
\chi_{pp}(q) = \frac{ \chi^{0}_{pp}(q)}{1-V_{pp}\chi^{0}_{pp}(q)} \; ,
                                                           \label{eq50}
\end{eqnarray}
where the irreducible susceptibility is defined as
$\chi^{0}_{pp}(q) = \frac{T}{4N}\sum_{k}
         (\phi_{d}(\vec{k})+\phi_{d}(\vec{q}-\vec{k}))^{2}
                                       G^{0}(q-k)G^{0}(k)$.
$G^{0}(k)$ is the noninteracting Green's function obtained 
from Eq.~(\ref{eq10}) with $J=0$.
Now the unknown vertex, $V_{pp}$, is determined 
by the sum rule Eq.~(\ref{eq40}).\cite{Vilk:1997}
Hence, an increase in   
the local spin-singlet amplitude evaluated 
in the interacting state over that in the noninteracting state 
leads to a nonvanishing positive value of $V_{pp}$, namely,
the enhancement of the correlation function.
This (non-perturbative sum rule) approach has been shown to be quite 
reliable\cite{Vilk:1997}
as long as short range correlations are concerned. 
In our calculations,                                    
the pseudogap appears when the spin-singlet
correlation length reaches about 1 lattice constant.
The self-energy due to the spin-singlet 
correlation function is given by\cite{Vilk:1997} 
\begin{eqnarray}
\Sigma_{pp}(k) &=& - \frac{1}{4}VV_{pp}
                \frac{T}{N}\sum_{q}
                                             \nonumber  \\
          & &  (\phi_{d}(\vec{k})+\phi_{d}(\vec{q}-\vec{k}))^{2}
                \chi_{pp}(q)G^{0}(q-k)
                                               \; ,
                                                           \label{eq70}
\end{eqnarray}
where $V=J$ from Eq.~(\ref{eq10}).

   First let us begin by showing the interaction-induced local  
spin-singlet (solid curve) amplitude (Fig.\ref{fig1}(b)) evaluated   
in the mean-field state of the $t-J$ Hamiltonian 
in a region where $s=0$ (or $T > T^{MF}_{c}$). 
Since $s=0$, the 
spin-singlet amplitude is entirely caused by short-range spin correlations
in the absence of pairing interactions.
Although in general a mean-field state is not accurate for  
strongly correlated electron systems,  
certain local and short-range {\em static} quantities such as
double occupancy and
nearest neighbor correlations
are reasonably well captured by the mean-field state particularly with AF order 
(See Ref.~\cite{Kyung:2000-3} for more details). 
In fact the interaction-induced local 
spin-singlet amplitude (Eq.~(\ref{eq40}))
is determined most crucially by these quantities.
Quite unexpectedly,
the local spin-singlet amplitude increases with decreasing doping  
despite the fact that the mean-field SC order $s$ is absent. 
The increase of local spin-singlet amplitude traces back to  
the growing short-range spin correlations  
with decreasing doping.\cite{Kyung:2000-3} 

   In Fig.~\ref{fig1}(a) we show the pseudogap temperature 
$T^{*}$ (filled diamonds) where the single particle spectral 
function $A(\vec{k},\omega)$ near $\vec{k}=(\pi,0)$ starts to 
be split into two peaks.
$T^{*}$ falls from
a high value onto the $T_{c}$ ($\le T_{c}^{MF}$) line  
instead of sharing a common 
line with $T_{c}$ in overdoped region.
It is not surprising to find that $T^{*}$
closely follows $T_{N}^{MF}=T^{0}$, because in our study the pseudogap is 
caused by induced local spin-singlet amplitude  
due to short-range spin correlations, 
which is reasonably well captured by the mean-field state with AF order.
When superconductivity is suppressed by setting $s=0$,
$T^{*}$ vanishes near $x_{c}$ where short-range spin correlations disappear.
All these features are at least qualitatively consistent with the 
findings by Tallon and Loram.\cite{Tallon:2001}
\begin{figure}
 \vbox to 6.5cm {\vss\hbox to -5.0cm
 {\hss\
       {\includegraphics{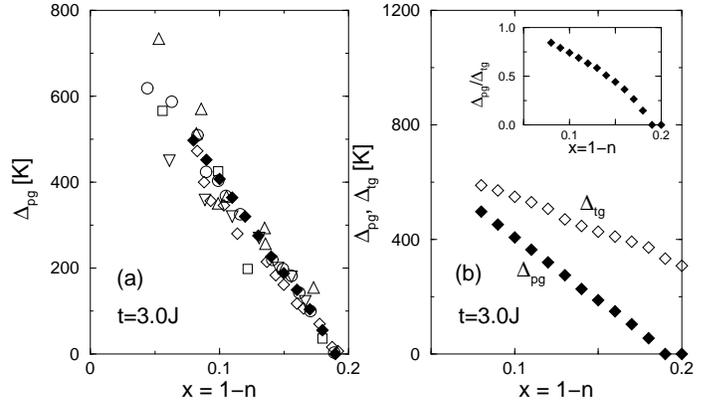}
       }
  \hss}
 }
\caption{(a) Pseudogap size
         denoted as filled diamonds at $T=0$ for $t/J=3.0$ and $J=125$ meV.
         The open circles, diamonds, squares, up-triangles, and down-triangles
         are the pseudogap size extracted from susceptibility,
         heat capacity, ARPES, NMR, and resistivity measurements, respectively,
         by Tallon and Loram.\protect\cite{Tallon:2001}
         (b) The total excitation gap $\Delta_{tg}$ (open diamonds) and
         pseudogap $\Delta_{pg}$ (filled diamonds)
         at $T=0$. The inset shows their relative ratio.}
\label{fig3}
\end{figure}

  Figure~\ref{fig3}(a) shows the pseudogap size $\Delta_{pg}$  
(filled diamonds)
by setting $s=0$ at $T=0$.
In the same figure the pseudogap energy extracted
from various experiments by Tallon and Loram\cite{Tallon:2001}
is also shown as empty symbols for comparison . 
$\Delta_{pg}$ vanishes near $x_{c}$,
suggesting the presence of a quantum critical point
at a critical doping.
The agreement between our results and experiments appears remarkable 
for such a simple approximation used in this paper.
The linear vanishing of $\Delta_{pg}$ near $x_{c}$ is closely
related to the corresponding behavior of the induced local
spin-singlet amplitude.

   The total excitation gap (or ARPES leading edge gap or SC gap) 
$\Delta_{tg}$ at $T=0$
is larger than $\Delta_{pg}$ due to the additional
contribution to the local spin-singlet amplitude from $s \ne 0$,
which is shown together with $\Delta_{pg}$
in Fig.~\ref{fig3}(b).
$\Delta_{pg}$, $\Delta_{tg}$ and 
their relative ratio $\Delta_{pg} / \Delta_{tg}$ 
are all monotonically decreasing functions of doping, 
as shown in the inset of Fig.~\ref{fig3}(b).
Since the SC order parameter vanishes at $T_{c}$
(at $T^{MF}_{c}$ in this paper), the SC gap below $T_{c}$ continuously
evolves into the normal state pseudogap above $T_{c}$ with
the same momentum dependence and magnitude.

   The pseudogap appears here only as the suppression of    
low frequency spectral weight   
in certain physical quantities 
which are obtained through $A(\vec{k},\omega)$ or its convolution 
with a relevant vertex.
It does not appear as a thermodynamic phase with broken symmetry.
The present scenario for the pseudogap predicts that a normal state pseudogap
is likely to appear when short-range spin correlations are well
established and are not masked by long-range (AF or SC) order.
The present results are robust to variations 
of $t/J=2.5-3.5$\cite{Comment40}
and small to moderate value of $t'$.

    In summary, we have shown that the 
induced local spin-singlet amplitude due to short-range spin correlations
causes a normal state pseudogap
even in the absence of pairing interactions.
$T^{*}$ falls from a high value onto the $T_{c}$ line and 
closely follows $T_{N}^{MF}$.
The calculated pseudogap size is in good agreement with 
experimental results.
It would be interesting to see how robust    
are the features found in this paper,   
when the no-double-occupancy constraint is strictly imposed 
on the $t-J$ model and an inhomogeneous solution is used.

   The author would like to thank Prof. A. M. Tremblay for numerous help and 
discussions throughout the work. He also thanks  
Profs. H. Ding, J. W. Loram, J. L. Tallon and T. Timusk
for useful discussions.
The present work was supported by a grant from the Natural Sciences and
Engineering Research Council (NSERC) of Canada and the Fonds pour la
formation de Chercheurs et d'Aide \`a la Recherche (FCAR) of the Qu\'ebec
government.

\begin{references}
\bibitem{Comment15}
The spin-singlet correlation function (with
$d$-wave symmetry) is more general than
the $d$-wave pair correlation function
in the sense that
the spin-singlet correlations can be also induced by
short-range spin correlations even in the absence of
pairing interactions.
\end{references}
\end{document}